\documentclass[10pt,showpacs,twocolumn]{revtex4}
\usepackage{graphicx}
\usepackage{amsmath}
\usepackage{amssymb}
\usepackage{}

\begin{document}
\title{Interference of high-order harmonics generated from molecules at different alignment angles}
\author{Meiyan Qin$^{1,2}$, Xiaosong Zhu$^{1,2}$, Yang Li$^{1,2}$,Qingbin Zhang$^{1,2}$, Pengfei Lan$^{1,2}$  \footnote{Corresponding author: pengfeilan@mail.hust.edu.cn}, and Peixiang Lu$^{1,2}$ \footnote{Corresponding author: lupeixiang@mail.hust.edu.cn}}

\affiliation{$^1$ School of Physics and Wuhan National Laboratory
for Optoelectronics, Huazhong University of
Science and Technology, Wuhan 430074, China\\
$^2$ Key Laboratory of Fundamental Physical Quantities Measurement of MOE, Wuhan 430074, China}
\date{\today}

\begin{abstract}
We theoretically investigate the interference effect of high-order
harmonics generated from molecules at different alignment angles.
It is shown that the interference of the harmonic emissions from
molecules aligned at different angles can significantly modulate
the spectra and result in the anomalous harmonic cutoffs observed
in a recent experiment {[ Nature Phys. 7, 822 (2011) ]}. The shift
of the spectral minimum position with decreasing the degree of
alignment is also explained by the interference effect of the
harmonic emissions.
\end{abstract}  \maketitle

\section{Introduction}
High-order harmonic generation (HHG) from molecules has attracted
a great deal of attention, due to the amazing application in
probing the molecular structure and electron dynamics with
attosecond and $\mathrm{\AA}$ngstr\"{o}m
resolutions\cite{lein,itatani,lan,torres,worner,zhu,haessler1,zhou}.
A rich set of new physical phenomena such as the spectral minimum
have been experimentally observed
\cite{kanai,vozzi1,liu,zhou1,zhou2,vozzi,lock}. In these
experiments, the target molecules were impulsively excited into a
rotational wave packet by a moderately intense pump pulse to
achieve the field-free alignment \cite{ortigoso,seideman}. And
then a more intense probe pulse at a certain time delay with
respect to the pump pulse was focused into these partially aligned
molecules to generate the high-order harmonics. The distribution
of the angle between the molecular axis and the polarization of
the probe pulse varies with the time delay. Therefore, to simulate
these pump-probe experiments, the alignment distribution of the
molecules at a certain time delay should be taken into account. In
previous theoretical works, two alternative methods were adopted
to take into account the partial alignment of molecules. In
Refs.\cite{liu,vozzi1,le}, a noncoherent superposition of the
harmonic intensities convolved with the angular distribution is
adopted to explain the experimental observations. While in Refs.
\cite{torres,vozzi,zhou1}, a coherent sum of the harmonic
emissions is employed to simulate the HHG from partially aligned
molecules.

The spectral minimum is one of the most important phenomena
observed in the molecular HHG. Both the interference of the
harmonic emissions from different nuclei in a molecule and the
interference of the harmonic emissions from multiple channels can
modulate the harmonic spectrum and induce a spectral minimum
\cite{lein1,smirnova}. With a detailed analysis of these minima,
the information about the molecular structure and electron
dynamics can be extracted \cite{lock,smirnova1}. It is shown that
the harmonic spectrum is also modulated due to the effect of the
partial alignment of molecules \cite{kanai,jin,rupenyan}, which is
inevitable in the actual experiment. Whereas how the interference
of the harmonic emissions from molecules at different alignment
angles affects the harmonic spectrum is seldom investigated.
Knowing this interference effect may provide a route to
quasi-phase-matching. Moreover, it helps us correctly extract the
information about the molecular structure and electron dynamics.

In this paper, we investigate the interference effect of the
harmonic emissions from molecules at different alignment angles.
By comparing the harmonic spectra obtained with the coherent and
noncoherent superpositions of the harmonic emissions with those
measured in experiment, it is confirmed that the harmonic
emissions from molecules at different alignment angles superpose
coherently. We perform a detailed analysis of the interference
effect at different time delays and with different degrees of
alignment. It is shown that the anomalous harmonic cutoff
phenomenon observed in a recent experiment \cite{vozzi} and the
shift of the spectral minimum position with decreasing the degree
of alignment can be explained by the interference effect of the
harmonic emissions from molecules at different alignment angles.
\section{Theoretical model}
To calculate the HHG spectrum for a fixed alignment, we use the
strong field approximation (SFA) model for molecules
\cite{lewen,qin}. Within the single active electron (SAE)
approximation, the time-dependent dipole velocity is given by
\begin{eqnarray}
\mathbf{v}_{dip}(t;\theta) & = & i\int^t_{-\infty}
dt'\left[\frac{\pi}{\zeta+i(t-t')/2}\right]^{\frac{3}{2}}\exp[-iS_{st}(t',t)]\nonumber\\
 & &\times\mathbf{F}(t')\cdot\mathbf{d}_{ion}\left[\mathbf{p}_{st}(t',t)+\mathbf{A}(t');\theta\right]
 \nonumber\\
 &
 &\times\mathbf{v}_{rec}^{*}\left[\mathbf{p}_{st}(t',t)+\mathbf{A}(t);\theta\right]+c.c..
 \end{eqnarray}
In this equation, $\zeta$ is a positive constant. $t'$ and $t$
correspond to the ionization and recombination time of the
electron, respectively. $\theta$ is the alignment angle between
the molecular axis and the polarization of the probe pulse.
$\mathbf{F}(t)$ refers to the electric field of the probe pulse,
and $\mathbf{A}(t)$ is its associated vector potential.
$\mathbf{p}_{st}$ and $S_{st}$ are the stationary momentum and the
quasi-classical action, which are given by
\begin{equation}
\mathbf{p}_{st}(t',t)=-\frac{1}{t-t'}\int^t_{t'}\mathbf{A}(t'')dt''
\end{equation}
and
\begin{equation}
S_{st}(t',t)=\int^t_{t'}(\frac{[\mathbf{p}_{st}+\mathbf{A}(t'')]^2}{2}+I_p)dt''
\end{equation}
with $I_p$ being the ionization energy of the state that the
electron is ionized from. Then the complex amplitude of the
high-order harmonics with a frequency $\omega_n$ is given by
\begin{equation}
\tilde{\mathbf{E}}(\omega_n,\theta)=\int e^{i\omega_n
t}\frac{d}{dt}\mathbf{v}_{dip}(t;\theta)dt.
\end{equation}

When the coherent superposition of the harmonic emissions is
employed to take into account the partial alignment of the
molecules, the spectrum at the delay $\tau$ with respect to the
pump pulse is given by
\begin{equation}
S(\omega_n;\tau)=|\int
E(\omega_n,\theta)exp[iP(\omega_n,\theta)]\rho(\theta;\tau)d\theta|^2.
\end{equation}
When the noncoherent superposition is adopted, the spectrum at the
delay $\tau$ is given by
\begin{equation}S(\omega_n; \tau)=\int
I(\omega_n,\theta)\rho(\theta;\tau)d\theta.
\end{equation}
In the two formulae,
$E(\omega_n,\theta)=|\tilde{\mathbf{E}}(\omega_n,\theta)|$,
$P(\omega_n,\theta)=\arg(\tilde{\mathbf{E}}(\omega_n,\theta))$,
and $I(\omega_n,\theta)=|\tilde{\mathbf{E}}(\omega_n,\theta)|^2$
are the amplitude, phase, and intensity of the high-order
harmonics generated from the molecule aligned at $\theta$.
$\rho(\theta;\tau)$ is the weighted angular distribution and is
given by
\begin{eqnarray}
\rho(\theta;\tau)&=&\sin\theta(1/Z)\sum_{J_i}Q(J_i)\nonumber\\
&\times&
\sum_{M_i=-J_i}^{J_i}\int|\Psi^{J_iM_i}(\theta,\phi;\tau)|^2
d\phi.
\end{eqnarray}
Here $Q(J_i)=exp(-BJ_i(J_i+1)/(k_BT))$ is the Boltzmann
distribution function of the initial field-free state
$|J_i,M_i\rangle$ at temperature $T$,
$Z=\sum^{J_{max}}_{J=0}(2J+1)Q(J)$ is the partition function,
$k_B$ and $B$ are the Boltzmann constant and the rotational
constant of the molecule, respectively.
$\Psi^{J_iM_i}(\theta,\phi;\tau)$ is the time-dependent rotational
wave packet excited from the initial state $|J_i,M_i\rangle$ by
the pump pulse, and is obtained by solving the time dependent
Schr\"{o}dinger equation (TDSE) within the rigid-rotor
approximation \cite{ortigoso,seideman}
\begin{equation}
i\frac{\partial\Psi(\theta,\phi;\tau)}{\partial
t}=[B\mathbf{J}^2-\frac{E_p(\tau)^2}{2}(\alpha_\parallel
\cos^2\theta+\alpha_\perp \sin^2\theta)]\Psi(\theta,\phi;\tau).
\end{equation}
In this equation, $\alpha_\parallel$ and $\alpha_\perp$ are the
anisotropic polarizabilities in parallel and perpendicular
directions with respect to the molecular axis, respectively. The
degree of alignment is characterized by the alignment parameter
$<\cos^2\theta>$, and is given by
\begin{equation}\label{eq:12}
\begin{split}
&<\cos^2\theta>(\tau) =(1/Z)\sum_{J_i}Q(J_i)\\
&\times\sum_{M_i=-J_i}^{J_i}\langle\Psi^{J_iM_i}(\theta,\phi;\tau)|\cos^2\theta|\Psi^{J_iM_i}(\theta,\phi;\tau)\rangle.
\end{split}
\end{equation}

\section{Result and discussion}
We choose the CO$_2$ molecule as an example for investigating the
interference effect of the harmonic emissions from molecules at
different alignment angles. In our simulation, a 100 fs (FWHM) 800
nm linearly polarized pump pulse with an intensity of
$4.0\times10^{13}$ $\mathrm{W/cm^2}$ is used to non-adiabatically
align the CO$_2$ molecule and the initial rotational temperature
is taken to be 40 K. For the CO$_2$ molecule, only even-J states
are populated in the ground state due to the spin statistics
\cite{miyazaki}. Hence, rotational states $|J,M\rangle$ with
$J=0,2,4...,24$ are included in our calculation. Then a delayed
probe pulse with the polarization parallel to that of the pump
pulse is used to generate high-order harmonics. The wavelength,
pulse duration, and intensity of the probe pulse are 1450 nm, 18
fs (FWHM), and $1.2\times10^{14}$ $\mathrm{W/cm^2}$, respectively.
These parameters of the pump and probe pulses are the same as
those used in the recent pump-probe experiment \cite{vozzi}. In
that work, the interference of harmonic emissions from multiple
channels is shown to be negligible when a few-cycle mid-IR (1450
nm) probe pulse is used. Therefore the HHG can be well described
within the SAE approximation.
\begin{figure}[t]
\setlength{\abovecaptionskip}{-0.1cm}
\setlength{\belowcaptionskip}{-0.15cm}
\centering\includegraphics[width=7.2cm]{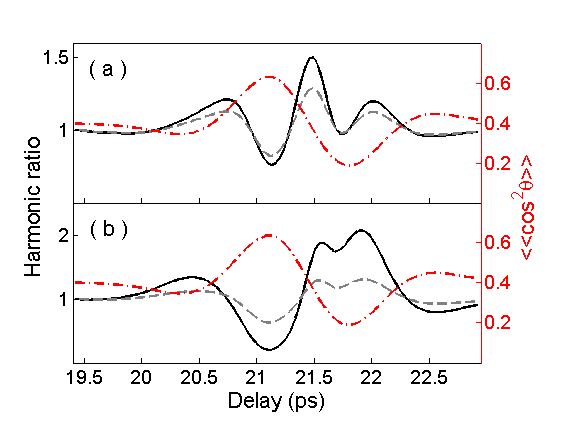} \caption{(Color
online) Ratio between harmonic signal generated in aligned and
unaligned CO$_2$ molecules for the harmonics at photon energies
35.72 eV (panel a) and 51.25 eV (panel b). The results obtained
with the coherent and noncoherent superposition methods are
presented by solid black and dashed grey lines respectively. The
time evolution of the alignment parameter $<\cos^2\theta>(\tau)$
is also displayed by the dash-dotted red line.}
\end{figure}

In our simulation, both the coherent and noncoherent
superpositions of the harmonic emissions are adopted to include
the partial alignment effect. We first calculate the time
evolution of the high-order harmonics at a fixed photon energy.
The time delay around the first half revival ($\tau=$21.1 ps) is
considered. In Fig. 1, the ratios between harmonic signals
generated from aligned and unaligned CO$_2$ molecules are
presented for the high harmonics at photon energies 35.72 eV
(panel a) and 51.25 eV (panel b). The solid black and dashed grey
lines show the results obtained with Eq. (5) and Eq. (6),
respectively. The time evolution of the alignment parameter
$<\cos^2\theta>(\tau)$ is also displayed by the dash-dotted red
line. As shown in Fig. 1, both the harmonic ratios obtained by the
coherent and noncoherent superposition present inverted modulation
with respect to the molecular alignment. These inverted
modulations of the harmonic ratios at 35.72 eV and 51.25 eV are
experimentally observed in Ref. \cite{kanai} and Ref.
\cite{vozzi1} respectively, and are explained by the two-center
interference effect. In Ref. \cite{vozzi1}, it is demonstrated
that with the noncoherent superposition method the inverted
modulation of the harmonic ratio can be satisfactorily reproduced.
Whereas our results show that not only the noncoherent but also
the coherent superpositions of the harmonic emissions can predict
the inverted modulations of the harmonic ratios at 35.72 eV and
51.25 eV.

\begin{figure}[b]
\centerline{\includegraphics[width=8.5cm]{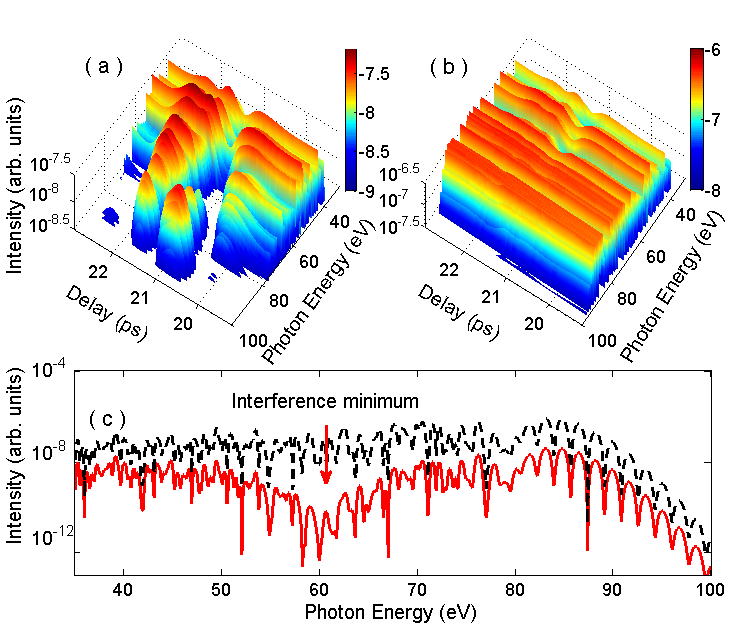}}
\setlength{\abovecaptionskip}{-0.1cm}
\setlength{\belowcaptionskip}{-0.15cm} \caption{(Color online)
Harmonic spectra as a function of the time delay obtained with
coherent (panel a) and noncoherent (panel b) superposition method.
(c) Harmonic spectra at the time delay 21.1 ps for the coherent
(solid red line) and noncoherent (dashed black line) superposition
cases.}
\end{figure}

We also calculate the whole spectrum as a function of the time
delay. The results are presented in Fig. 2. As shown in Fig. 2(a),
the harmonic spectra obtained by employing the coherent
superposition agree well with the observations of the recent
pump-probe experiment \cite{vozzi}. In detail, obvious spectral
minima are observed at the delays around 21.1 ps. Furthermore, the
spectral cutoffs at the delays around 21.1 ps and 21.81 ps appear
at 84.7 eV. While in the harmonic spectra at other delays the
cutoffs are observed at much lower photon energies, which can not
be explained by the cutoff-law $I_p+3.17U_p$. This anomalous
cutoff phenomenon is described as the cutoff recession in Ref.
\cite{vozzi}. As for the noncoherent superposition case shown in
Fig. 2(b), however, both the interference minima and the cutoff
recession phenomenon are missing. For clarity, the harmonic
spectra at the delay time 21.1 ps are presented in Fig. 2(c) for
the coherent (the solid red line) and noncoherent (the dashed
black line) superposition cases. As indicated by the red arrow, an
obvious minimum occurs only in the spectrum obtained by employing
the coherent superposition. Moreover, the spectral minimum
position (60.3 eV) is the same as that measured in Ref.
\cite{vozzi}. Therefore, only the coherent superposition of the
harmonic emissions from molecules aligned at different angles can
fully describe the HHG from partially aligned molecules. Although
some features of the HHG can be reproduced by the noncoherent
superposition of the harmonic emissions as shown in Fig. 1 and
also shown in Ref. \cite{vozzi1}, the more accurate description of
the HHG from partially aligned molecules is the coherent
superposition. Our results confirm that the harmonic emissions
from molecules at different alignment angles superpose coherently.

To provide insights into the anomalous harmonic cutoff phenomenon
observed in experiment \cite{vozzi}, we perform an analysis of the
angular-dependent amplitudes and phases of the harmonics near the
cutoff in perfect alignment case. In Fig. 3(a), the amplitude
($E(\omega_n,\theta)$) and phase ($P(\omega_n,\theta)$) of the
harmonic emission at photon energy 77.5 eV, where the harmonic
cutoff recession occurs, are presented. As shown in Fig. 3(a), the
harmonic emissions near the cutoff exhibit two-center interference
minima around $50^\circ$, which is accompanied by a phase jump of
$\pi$. Thus, the harmonic emissions are divided by the $50^\circ$
angle into two parts that have $\pi$ phase difference. For the
first part $\theta<50^\circ$, the corresponding harmonic phases
are around $0.5\pi$, while for the second part $\theta>50^\circ$,
the phases are around $-0.5\pi$. As a result, when coherently
superposing the harmonic emissions at different alignment angles,
the interference of the harmonic emissions will modulate the
observed spectra, depending on the corresponding angular
distribution. In Figs. 3(b-d), the polar plot of
$\rho(\theta;\tau)$ at three typical delays 21.1 ps, 21.81 ps, and
22.85 ps are presented, respectively. At the first half revival
delay 21.1 ps, the CO$_2$ molecules are mostly aligned along the
laser polarization direction ($\theta=0^\circ$). The corresponding
weighted angular distribution maximizes at $24^\circ$, as shown in
Fig. 3(b). Thus at delays around 21.1 ps the harmonics near the
cutoff are determined by the constructive interference of the
harmonic emissions from molecules aligned at the angles
$\theta<50^\circ$. At delays around 21.81 ps, the CO$_2$ molecules
are mostly aligned
\begin{figure}[t]
\centering\includegraphics[width=8.0cm]{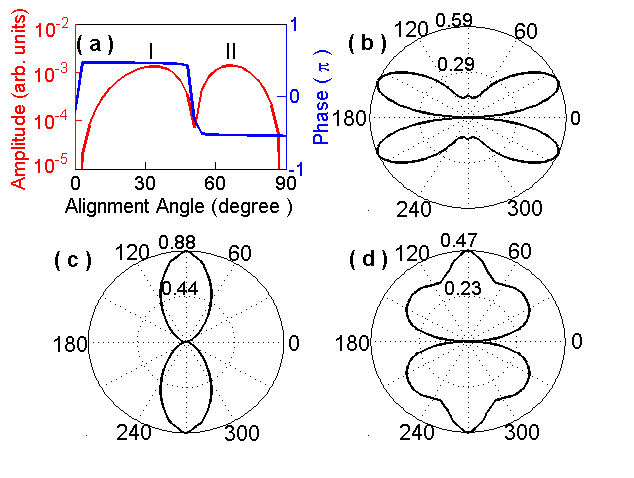}
\setlength{\abovecaptionskip}{-0.1cm}
\setlength{\belowcaptionskip}{-0.15cm} \caption{(Color online) (a)
The angular-dependent amplitude (solid red line) and phase (solid
blue line) of the harmonic emission at photon energy 77.5 eV in
perfect alignment case. (b-d) The polar plots of the angular
distributions $\rho(\theta;\tau)$ at the delays 21.1 ps (b), 21.81
ps (c) and 22.85 ps (d).}
\end{figure}
perpendicular to the laser polarization ($\theta=90^\circ$). The
corresponding weighted angular distribution maximizes at
$90^\circ$, as shown in Fig. 3(c). Thus the harmonics near the
cutoff are determined by the constructive interference of the
harmonic emissions from molecules at the alignment angles
$\theta>50^\circ$. Correspondingly, the spectral cutoffs at delays
around 21.1 ps and 21.81 ps appear at the photon energy 84.7 eV,
which agrees with the cutoff-law $I_p+3.17U_p$. While at other
delays (for example at 22.85 ps as shown in Fig. 3(d)), the
molecular alignment is quasi-random. The emissions at the
alignment angles below and above $50^\circ$ contribute almost
equally to the high-order harmonics near the cutoff, i.e. the
contributions of the two parts with $\pi$ phase difference are
comparable. Hence the harmonics near the cutoff are significantly
suppressed due to the destructive interference. As a result, the
harmonic spectra present cutoff recessions with respect to those
around 21.1 ps and 21.81 ps. From the above analysis, one can see
that the anomalous cutoff phenomenon results from the interference
of the emissions from molecules aligned at different angles.

\begin{figure}[htb]
\centering\includegraphics[width=8.0cm]{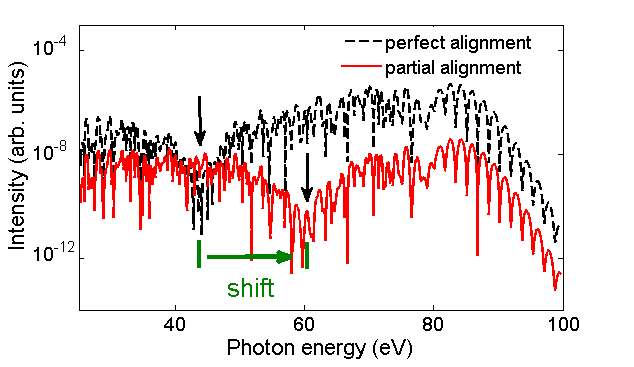}
\setlength{\abovecaptionskip}{-0.1cm}
\setlength{\belowcaptionskip}{-0.15cm} \caption{(Color online) The
harmonic spectra generated from CO$_2$ molecules in perfect
alignment at $24^\circ$ (dashed black line) and in partial
alignment at the delay 21.1 ps (solid red line).}
\end{figure}

When the destructive interference of the harmonic emissions from
molecules at different alignment angles occurs in the cutoff
region, the cutoff recession appears. While if the destructive
interference occurs at lower photon energies (i.e., in the plateau
region), a well-defined spectral minimum will be observed. For
high-order harmonics at a low photon energy, the associated phase
jump by $\sim\pi$ occurs at a small angle. Accordingly, the
destructive interference between the harmonic emissions at
different alignment angles appears at the delays (such as 21.1
ps), when the molecules are mostly aligned at small angles. As a
result, well-defined spectral minima are observed at the delays
around 21.1 ps, as presented in Fig. 2(a). In the following, we
compare the minimum positions in the harmonic spectra generated
from perfectly and partially aligned molecules. As shown in Fig.
3(b), the angular distribution at the delay 21.1 ps for the CO$_2$
molecules maximizes at $24^\circ$. Therefore, we use the harmonic
spectrum from CO$_2$ molecules perfectly aligned at $24^\circ$ as
a reference. In Fig. 4, both the harmonic spectra generated from
perfectly (dashed black line) and partially (solid red line)
aligned molecules are presented. The horizontal axis represents
the photon energy. As shown in Fig. 4, the interference minimum in
the harmonic spectrum for the perfect alignment is observed at
photon energy 43.8 eV, which agrees with the two-center
interference model with the dispersion relation
$n\hbar\omega=E_k+I_p$. While after taking into account the
partial alignment of molecules, the minimum appears at a higher
energy 60.3 eV. Hence, the interference of the harmonic emissions
at different alignment angles significantly shifts the spectral
minimum position.

\begin{figure}[htb]
\centering\includegraphics[width=9.0cm]{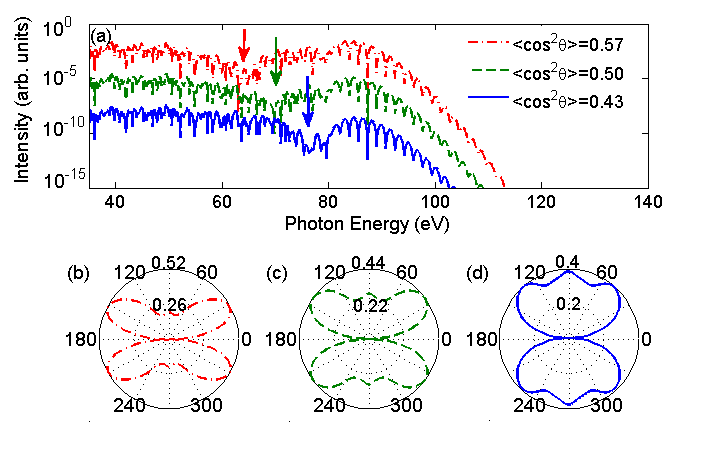}
\setlength{\abovecaptionskip}{-0.1cm}
\setlength{\belowcaptionskip}{-0.15cm} \caption{(Color online) (a)
Harmonic spectra at the first half revival with $<\cos^2\theta>$
as 0.57 (dash-dotted red line), 0.50 (dashed green line) and 0.43
(solid blue line). The harmonic spectra for $<\cos^2\theta>$ as
0.57 and 0.50 are vertically shifted for clarity. (b-d) Polar plot
of the corresponding weighted alignment distributions for
$<\cos^2\theta>$ as 0.57 (b), 0.50 (c) and 0.43 (d).}
\end{figure}

Next, we investigate how the spectral minimum position is affected
by the interference of the harmonic emissions from molecules at
different alignment angles. In Fig. 5(a), the harmonic spectra at
the first half revival are presented for $<\cos^2\theta>$ as 0.57
(dash-dotted red line), 0.50 (dashed green line), and 0.43 (solid
blue line). These three degrees of alignment are achieved by using
the pump pulse with intensities $3.0\times10^{13}\
\mathrm{W/cm^2}$, $2.0\times10^{13}\ \mathrm{W/cm^2}$, and
$1.0\times10^{13}\ \mathrm{W/cm^2}$, respectively. As shown in
Fig. 5(a), the minimum position shifts to higher photon energies
when the degree of alignment decreases. The same phenomenon has
been observed experimentally in Ref. \cite{rupenyan}. To explain
this phenomenon, the corresponding weighted alignment
distributions are presented in Figs. 5(b-d) for these three
degrees of alignment. As shown in Figs. 5(b-d), the angle where
the weighted alignment distribution maximizes shifts to a large
angle with decreasing the degree of alignment. Simultaneously, the
contributions from large alignment angles gradually increase. As
discussed in Fig. (3), the critical angle, where a phase jump
occurs, divides the harmonics into two parts with a $\pi$ phase
difference. When the contributions from the two parts are
comparable, the most significant destructive interference of the
harmonic emissions from different alignments (i.e. a spectral
minimum) will occur at the corresponding high-order harmonics. In
the case with a low value of $<\cos^2\theta>$, the critical angle
that associates with a phase jump of $\pi$ and divides the
harmonic emissions into two comparable parts shifts to a large
angle. According to the two-center interference model, the
high-order harmonics with the phase jump at a larger angle has a
higher photon energy. As a result, the spectral minimum will shift
to a higher energy with the degree of alignment decreasing.

\section{Conclusion}
In conclusion, we investigate the interference effect of the
harmonic emissions from molecules at different alignment angles.
By comparing the harmonic spectra obtained in our simulation with
those measured in experiment, it is confirmed that the harmonic
emissions from molecules at different alignment angles superpose
coherently. We perform a detailed analysis of the interference
effect of the high-order harmonics from partially aligned
molecules at different time delays and with different degrees of
alignment. It is shown that the interference of the harmonic
emissions from molecules aligned at different angles can
significantly modulate the harmonic spectrum and result in the
cutoff recession phenomenon observed in a recent experiment
\cite{vozzi}. The shift of the spectral minimum position with
decreasing the degree of alignment is also explained by the
interference effect of the harmonic emissions.
\section{ACKNOWLEDGMENTS}
This work was supported by the NNSF of China under Grants No.
11234004 and 60925021, the 973 Program of China under Grant No.
2011CB808103 and the Doctoral fund of Ministry of Education of
China under Grant No. 20100142110047.


\begin{thebibliography}{99}
\bibitem{lein} M. Lein, N. Hay, R. Velotta, J. P. Marangos, and P.
L. Knight, Phys. Rev. Lett. {\bf 88}, 183903 (2002).

\bibitem{itatani} J. Itatani, J. Levesque, D. Zeidler, H. Niikura,
H. P\'{e}pin, J. C. Kieffer, P. B. Corkum and D. M. Villeneuve,
Nature {\bf 432}, 867 (2004).

\bibitem{lan} P. Lan {\it et al.}, Phys.
Rev. A {\bf 76}, 011402(R) (2007); W. Hong, {\it et al.} J. Opt.
Soc. Am. B {\bf 25} 1684-1689 (2008); J. Luo {\it et al.} J. Phys.
B {\bf 46} 145602 (2013).

\bibitem{torres} R. Torres, N. Kajumba, J. G. Underwood, J. S.
Robinson, S. Baker, J. W. G. Tisch, R. deNalda, W. A. Bryan, R.
Velotta, C. Altucci, I. C. E. Turcu, and J. P. Marangos, Phys.
Rev. Lett. {\bf 98}, 203007 (2007).

\bibitem{worner} H. J. W\"{o}rner, J. B. Bertrand, D. V.
Kartashov, P. B. Corkum and D. M. Villeneuve, Nature {\bf 466},
604 (2010).

\bibitem{zhu} X. Zhu, Q. Zhang, W. Hong, P. Lan, and P. Lu,  Opt. Express {\bf 19}, 436 (2010).

\bibitem{haessler1} S. Haessler, J. Caillat and P. Sali\`{e}res, J.
Phys. B: At. Mol. Opt. Phys. {\bf 44}, 203001 (2011).

\bibitem{zhou} Y. Zhou {\it et al.} Phys. Rev. Lett. {\bf
109}, 053004 (2012); Q. Liao {\it et al.} New J. Phys. {\bf 14}
013001 (2012); C. Huang {\it et al.} Opt. Express {\bf 21} 11382
(2013).

\bibitem{kanai} T. Kanai, S. Minemoto and H. Sakai, Nature {\bf 435},
470 (2005).

\bibitem{vozzi} C. Vozzi, M. Negro, F. Calegari, G. Sansone, M.
Nisoli, S. De Silvestri and S. Stagira, Nature Phys. {\bf 7}, 822
(2011).

\bibitem{zhou1} X. Zhou, R. Lock, W. Li, N. Wagner, M. M. Murnane,
and H. C. Kapteyn, Phys. Rev. Lett. {\bf 100}, 073902 (2008).

\bibitem{liu} P. Liu, P. Yu, Z. Zeng, H. Xiong, X. Ge, R. Li, and
Z. Xu, Phys. Rev. A {\bf 78}, 015802 (2008).

\bibitem{vozzi1} C. Vozzi, F. Calegari, E. Benedetti, J. P.
Caumes, G. Sansone, S. Stagira, and M. Nisoli, R. Torres, E.
Heesel, N. Kajumba, and J. P. Marangos, C. Altucci, R. Velotta,
Phys. Rev. Lett. {\bf 95}, 153902 (2005).

\bibitem{zhou2} X. Zhou, R. Lock, N. Wagner, W. Li, H. C. Kapteyn, and M. M. Murnane,
Phys. Rev. Lett. {\bf 102}, 073902 (2009).

\bibitem{lock} R. M. Lock, S. Ramakrishna, X. Zhou, H. C. Kapteyn,
M. M. Murnane, and T. Seideman, Phys. Rev. Lett. {\bf 108}, 133901
(2012).

\bibitem{ortigoso} J. Ortigoso and M. Rodriguez, M. Gupta and B. Friedrich, J. Chem.
Phys. {\bf 110}, 3870 (1999).

\bibitem{seideman} T. Seideman, Phys. Rev. Lett. {\bf 83}, 4971 (1999).

\bibitem{le} A. T. Le, X. M. Tong, and C. D. Lin, Phys. Rev. A {\bf 73}, 041402(R) (2006).

\bibitem{lein1} M. Lein, N. Hay, R. Velotta, J. P. Marangos, and P.
L. Knight, Phys. Rev. A. {\bf 66}, 023805 (2002).

\bibitem{smirnova} O. Smirnova, S. Patchkovskii, Y. Mairesse, N. Dudovich, D. Villeneuve,
P. Corkum, and M. Yu. Ivanov, Phys. Rev. Lett. {\bf 102}, 063601
(2009).

\bibitem{smirnova1} O. Smirnova, Y. Mairesse, S. Patchkovskii, N. Dudovich, D. Villeneuve,
P. Corkum, and M. Yu. Ivanov, Nature {\bf 460}, 972 (2009).

\bibitem{jin} C. Jin, A. T. Le, and C. D. Lin, Phys. Rev. A. {\bf 83}, 053409 (2011).

\bibitem{rupenyan} A. Rupenyan, P. M. Kraus, J. Schneider, and H. J. W\"{o}rner, Phys. Rev. A {\bf 87}, 031401(R) (2013).

\bibitem{lewen} M. Lewenstein, Ph. Balcou, M. Yu. Ivanov, A. L'Huillier,
and P. B. Corkum, Phys. Rev. A {\bf 49,} 2117 (1994).

\bibitem{qin} M. Qin, X. Zhu, Q. Zhang, and P. Lu, Opt. Letters {\bf 37}, 5208 (2012).

\bibitem{miyazaki} K. Miyazaki, M. Kaku, G. Miyaji, A. Abdurrouf,
and F. H. M. Faisal, Phys. Rev. Lett. {\bf 95}, 243903 (2005).

\end{thebibliography}
\end{document}